\documentclass[twocolumn,showpacs,preprintnumbers,amsmath,amssymb,superscriptaddress,prb]{revtex4}
\usepackage{epsfig}
\usepackage{graphicx}

\newcommand{\E}{{\rm e}}

\newcommand{\I}{{\rm i}}

\newcommand{\D}{{\rm d}}

\newcommand{\beq}[1]{
\begin{equation}
\label{e#1} }

\newcommand{\eeq}{
\end{equation}
}

\begin{document}

\title{Mn incorporation in as-grown and annealed (Ga,Mn)As layers studied by x-ray diffraction and
standing-wave fluorescence}
\author{V. Hol\'{y}}
\author{Z. Mat\v{e}j}
\affiliation{Charles University, Faculty of Mathematics and
Physics, Department of Electronic Structures, Ke Karlovu 5, 121 16
Prague 2, Czech Republic}
\author{O. Pacherov\'a}
\author{V. Nov\'ak}
\author{M. Cukr}
\author{K. Olejn\'{\i}k}
\affiliation{Institute of Physics ASCR, Cukrovarnick\'a 10, 162 53
Praha 6, Czech Republic}
\author{T. Jungwirth}
\affiliation{Institute of Physics ASCR, Cukrovarnick\'a 10, 162 53
Praha 6, Czech Republic} \affiliation{School of Physics and
Astronomy, University of Nottingham, Nottingham NG7 2RD, UK}
\date{\today}
\begin{abstract}
A combination of high-resolution x-ray diffraction and a new
technique of x-ray standing wave fluorescence at grazing incidence
is employed to study the structure of (Ga,Mn)As diluted magnetic
semiconductor and its changes during post-growth annealing steps.
We find that the film is formed by a uniform, single crystallographic phase epilayer
covered by a thin surface layer with enhanced Mn concentration due
to Mn atoms at random non-crystallographic positions. In the
epilayer, Mn incorporated at interstitial position has a dominant
effect on lattice expansion as compared to substitutional Mn. The
expansion coefficient of interstitial Mn estimated from our data
is consistent with theory predictions. The concentration of
interstitial Mn and the corresponding lattice expansion of the
epilayer are reduced by annealing, accompanied by an increase of
the density of randomly distributed Mn atoms in the disordered
surface layer. Substitutional Mn atoms remain stable during the
low-temperature annealing.
\end{abstract}

\maketitle

\section{Introduction}

\begin{figure}
\hspace*{-1.5cm}\epsfig{width=5cm,angle=-90,file=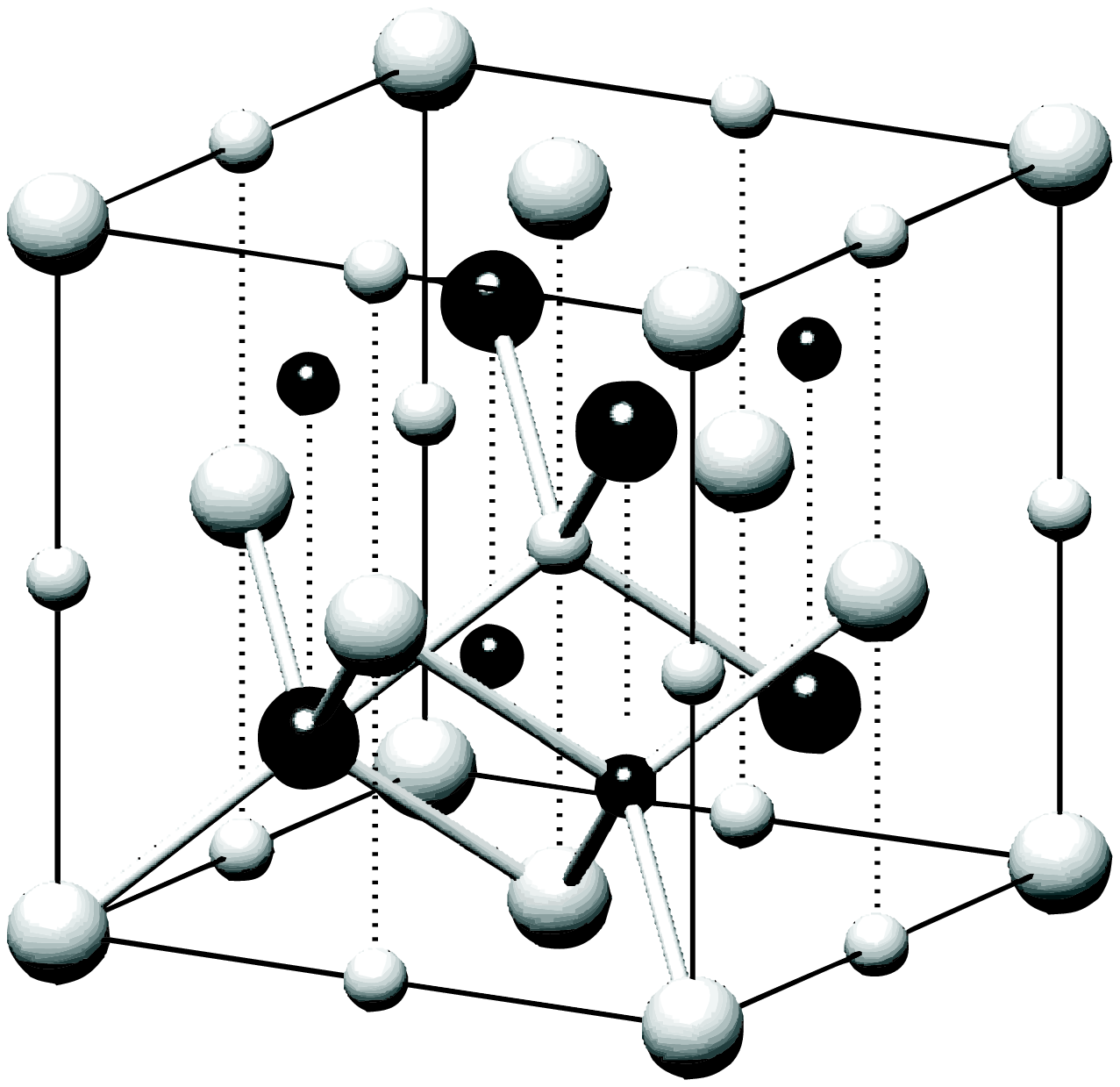}

\vspace*{-4cm}
\hspace*{6.8cm}\epsfig{width=1.8cm,angle=0,file=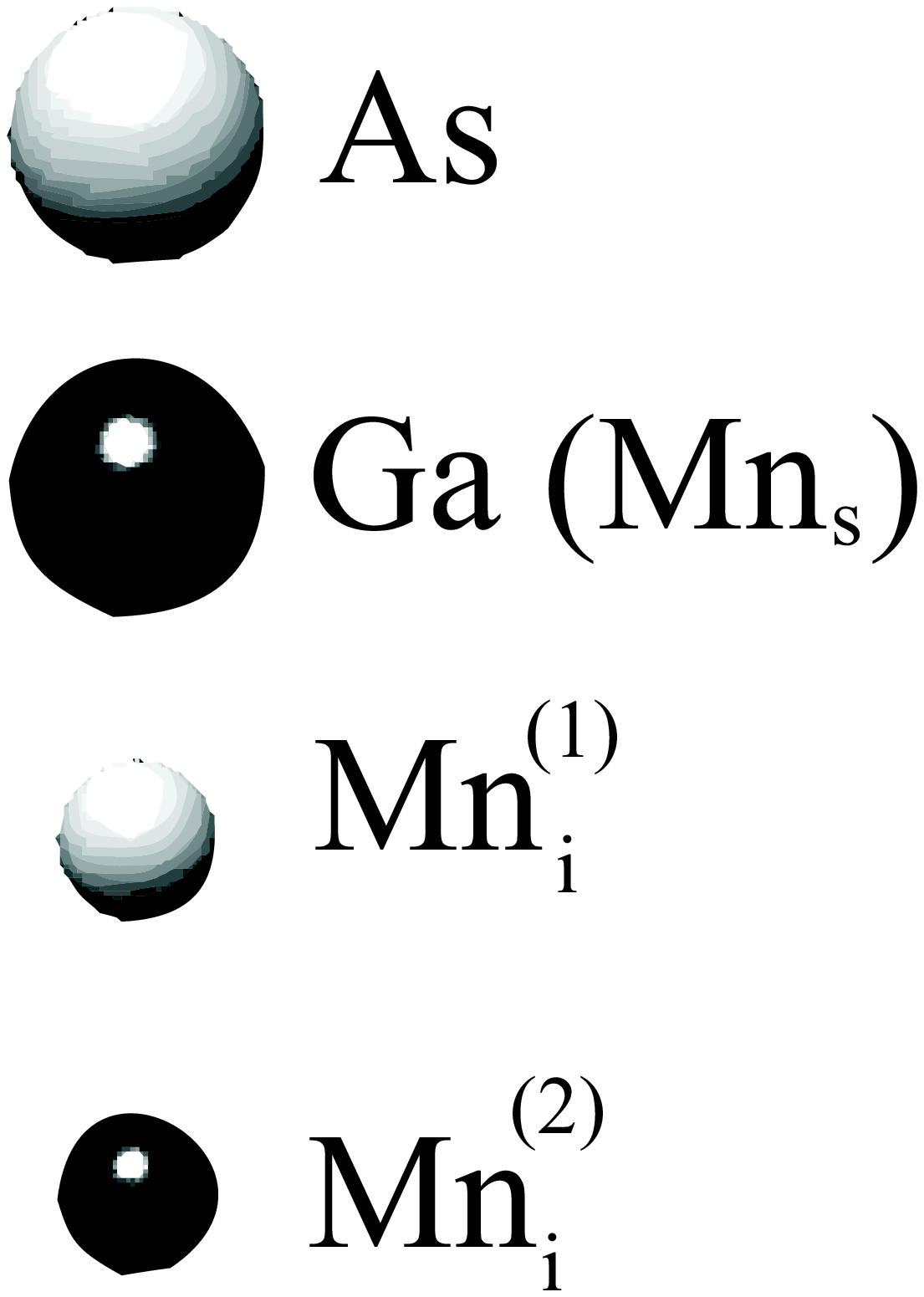}

\vspace*{1.5cm}
\caption{The elementary unit
cell of GaAs with possible positions of Mn atoms. The interstitial
atoms Mn$_{\I}^{(2)}$ have the same lateral coordinates as the
substitutional Mn atoms.}
\label{Figure1}
\end{figure}

Mn is incorporated in ferromagnetic (Ga,Mn)As semiconductors in
substitutional Mn$_{\rm Ga}$ positions and fraction of Mn atoms
can also occupy interstitial sites surrounded by four Ga
(Mn$_{\I}^{(1)}$) or four As (Mn$_{\I}^{(2)}$) nearest neighbors,
as illustrated in Fig.~\ref{Figure1}.
Substitutional Mn$_{\rm Ga}$ ions provide local magnetic moments
and holes that mediate long-range ferromagnetic coupling between
the local moments.\cite{Matsukura:2002_a,Jungwirth:2006_a} The
less energetically favorable Mn$_{\I}$
donors\cite{Yu:2002_a,Maca:2002_a} occur due to the tendency to
self-compensation in the highly Mn-doped GaAs host with nearly
covalent crystal bonding. Mn$_{\I}$ impurities  are detrimental to
ferromagnetism since they act as charge and moment compensating
defects.\cite{Blinowski:2003_a,Masek:2003_b,Edmonds:2004_a,Jungwirth:2005_b,Jungwirth:2005_a}

The concentration of Mn$_{\I}$ in (Ga,Mn)As epilayers can be
significantly reduced by low-temperature ($\sim 200^{\circ}$C)
post-growth annealing  which leads to a dramatic improvement of
ferromagnetic properties of the
epilayers.\cite{Yu:2002_a,Edmonds:2002_b,Chiba:2003_b,Ku:2003_a,Stone:2003_a}
For example, in the material with 9\% nominal Mn doping which
holds current record Curie temperature, the
paramagnetic-to-ferromagnetic transition was moved from 95~K in
the as-grown material up to 173~K after
annealing.\cite{Jungwirth:2005_b} Detailed analysis of Mn
incorporation in (Ga,Mn)As  and of the process of Mn$_{\I}$
removal by annealing has therefore been one of the key topics in
materials research of these dilute moment ferromagnetic
semiconductors.

Direct experimental evidence for the presence of interstitial Mn
impurities in as-grown (Ga,Mn)As epilayers and for their reduced
density after post-growth annealing was given by combined
channeling Rutherford backscattering and particle induced x-ray
emission measurements.\cite{Yu:2002_a} The technique can
distinguish between Mn$_{\I}$ and Mn$_{\rm Ga}$ by counting the
relative number of exposed Mn atoms and the ones shadowed by
lattice site host atoms at different channeling angles. Mn
incorporation at different crystallographic positions was also
studied by x-ray absorption\cite{Ishiwata:2002_a}, and by
high-resolution diffraction and structure factor
measurements.\cite{Glas:2004_a,Frymark:2005_a} The extraction of
partial concentrations of different types of Mn impurities in the
lattice by these techniques is less straightforward as the data
are sensitive not only to the crystallographic positions of Mn but
also to local lattice distortions on neighboring sites. X-ray
diffraction measurement of the (Ga,Mn)As epilayer lattice
parameter is another indirect technique that has been used to
monitor the decrease of Mn$_{\I}$ concentration in the epilayer
after
annealing.\cite{Sadowski:2004_a,Kuryliszyn-Kudelska:2004_a,Zhao:2005_a}
Here the interpretation has relied on the assumption that
Mn$_{\I}$ are the only mobile impurities during the annealing
process,\cite{Erwin:2002_a,Masek:2003_b,Bliss:1992_a} and
used\cite{Sadowski:2004_a} theoretical values\cite{Masek:2003_a}
for the defect expansion coefficients in (Ga,Mn)As or
assumed\cite{Zhao:2005_a} specific scenarios for charge
compensation that fulfill the overall charge neutrality condition.

A series of experimental works, including resistance monitored
annealing combined with Auger spectroscopy,\cite{Edmonds:2004_a}
depth-profiling by x-ray reflectometry,\cite{Kirby:2006_a} and
surface capping
experiments,\cite{Stone:2003_a,Adell:2004_b,Malfait:2004_a,Kirby:2004_a}
have focused on the mechanism causing the decrease of Mn$_{\I}$
concentration in annealed (Ga,Mn)As epilayers. The studies suggest
that Mn$_{\I}$ impurities outdiffuse to the layer surface and are
passivated by oxygen when annealed in air or by forming MnAs in
case of (Ga,Mn)As films overgrown by As capping layers.

In this paper we report on high resolution x-ray
diffraction\cite{Pietsch:2003_a} (XRD) experiments  and
measurements of x-ray standing wave\cite{Zegenhagen:1993_a} (XSW)
fluorescence at grazing incidence in as-grown and annealed
(Ga,Mn)As films. The former technique allows us to monitor changes
in the abundance of different types of impurities in the ordered
part of the epilayer during annealing steps by determining
individual contributions from the impurities to the lattice
parameter. XSW at grazing incidence is a new experimental
technique which we developed for detecting lattice-site resolved
Mn fluorescence from pseudomorphic (Ga,Mn)As thin films. By small
variations in the incidence angle of the primary beam we can tune the
exposure from covering the entire epilayer to only a few nanometer thin surface
layer. Data collected by the two complementary techniques are
interpreted directly with no \emph{a priori} assumptions on the
behavior of Mn in as-grown systems and during annealing. They are
used to determine crystal phases in the (Ga,Mn)As layers, to
compare lattice expansion contributions from Mn$_{\rm Ga}$ and
Mn$_{\I}$, to give an estimate for the Mn$_{\I}$ lattice expansion
coefficient, to compare stabilities of Mn$_{\rm Ga}$ and Mn$_{\I}$
atoms during annealing, and to directly monitor Mn$_{\I}$
outdiffusion from the (Ga,Mn)As epilayer to the surface and to
characterize the surface Mn incorporation.

The paper is organized as follows. In Section~\ref{experiment} we
briefly introduce the sample growth and characterization and
experimental set-ups. Because of the novelty of the grazing
incidence XSW technique we derive in Section~\ref{theory}
expressions for the fluorescence intensity in this geometry and
show typical numerical spectra used for fitting the experimental
data. Results of XSW and XRD measurements and their comparison
with theory are presented in Section~\ref{results}. In section
\ref{discussion} we discuss our results in the context of previous and potential future
measurements, and we
conclude in Section~\ref{conclusions} with a brief summary of our
main findings.

\section{Experiment}
\label{experiment}

Four samples were studied from a wafer grown by molecular-beam
epitaxy (MBE) on a (001) GaAs semi-insulating substrate. The MBE
deposition started with a 320~nm thick high-temperature GaAs
buffer followed by a 5~nm low-temperature GaAs layer and, finally,
a low-temperature (Ga,Mn)As epilayer was grown with a nominal
thickness of 100~nm. The substrate temperature during the
low-temperature deposition was 240$^\circ$C as measured by a floating thermocouple. Mn doping of $3.5\pm
0.5$\% was estimated from the ratio of the growth rates measured
by the RHEED oscillations before and after opening the Mn cell.
The as-grown structure (sample \#1) shows metallic conductivity
with the room temperature conductivity of
300~$\Omega^{-1}$cm$^{-1}$. The Curie temperature of 55~K was
determined by SQUID magnetization measurements. Samples \#2, \#3,
and \#4 were annealed in air at 200$^\circ$C for 1, 2, and 4
hours, respectively. Corresponding Curie temperatures are 65~K,
68~K, and 72~K.

Room temperature XRD measurements of the thickness of (Ga,Mn)As
epilayers and of the contributions to the lattice parameter from
different types of impurities were performed for coplanar
symmetric 002 and 004 diffractions. We used a commercial
high-resolution diffractometer equipped with conventional
CuK$\alpha$ x-ray tube, parabolic x-ray mirror, Bartels-type 4x220
Ge monochromator, and 3x220 Ge analyzing crystal.

XSW fluorescence experiments were carried out on the ID32 beamline
at the European Synchrotron Radiation Facility in Grenoble, using
the energy of the primary x-ray beam of 10~keV. The technique is
based on the interference of the primary x-ray beam with the
diffracted beam which creates a standing wave in the diffracting
crystal volume. The period of the standing-wave pattern equals the
distance of the diffracting net planes and the position of the
antinodes of the standing wave can be sensitively tuned by
changing the direction of the primary
beam.\cite{Zegenhagen:1993_a,Vartanyants:2001_a} The presence of Mn atoms in 
different lattice positions  was detected by measuring the fluorescent
MnK$\alpha$ radiation. The solid-state detector was placed close
to the sample surface, collecting the fluorescence signal from a wide
solid angle, yet avoiding direct illumination of the
detector by the primary and diffracted beams.

\begin{figure}
\epsfig{width=6cm,file=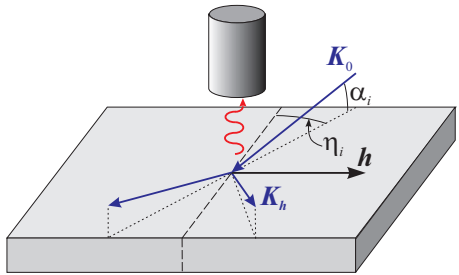}
\caption{Sketch of the grazing-incidence geometry. The wavy line
is the fluorescence radiation, the dashed line denotes the
diffracting crystallographic plane perpendicular to the sample
surface.}
\label{Figure2}
\end{figure}

The XSW experimental set-up is schematically illustrated in
Fig.~\ref{Figure2}. For in-plane 400, 220, 200, and 420
diffractions, and for various incidence angles $\alpha_i$ near the
critical angle $\alpha_c$ of total external reflection, we
measured the dependence of the fluorescence signal $\Phi$ on the
azimuthal deviation $\delta\eta_i$ of the primary x-ray beam from
the diffraction position. 
Recall that (Ga,Mn)As
epitaxial layers grown on GaAs buffer have a pseudomorphic
structure, i.e., their in-plane lattice parameter is the same as
in the substrate, however, their vertical lattice parameters are different.
Therefore, the XSW experiment has to be carried out in the grazing incidence ($\alpha_i\sim \alpha_c$)
geometry.
In this arrangement the diffracting net planes and the antinode planes of the standing wave are perpendicular to the
surface allowing only for the determination of the in-plane coordinates of Mn atoms.

To our knowledge this technique has not yet been employed and we,
therefore, included the following theory section which details the
procedures for analyzing grazing incidence XSW data.

\section{Theory}
\label{theory}

In the calculation of the fluorescence intensity we assume that
the intensity of the fluorescence radiation emitted from a given
atom is proportional to the magnitude of the Poynting vector
\beq{1}
{\bf S}=\frac{1}{2}{\rm Re}({\bf E}\times{\bf H}^*)
\eeq
of the wave field inside the (Ga,Mn)As layer in the position of a
given Mn atom. For the calculation of the electric and magnetic
intensities ${\bf E}$ and ${\bf H}$ we use the two-beam
approximation of dynamical diffraction
theory.\cite{Authier:2001_a} In this approximation, the wave-field
inside the layer can be written as a superposition of four
transmitted waves (${\bf E}_{0n},\ n=1,\ldots,4$) and four
diffracted waves (${\bf E}_{{\bf h}n}\equiv c_n {\bf E}_{0n},\
n=1,\ldots,4$) with wave vectors ${\bf k}_{0n}$ and ${\bf k}_{{\bf
h}n}={\bf k}_{0n}+{\bf h}$, respectively; ${\bf h}$ is the
diffraction vector which is assumed to be parallel to the surface.
In the following, we restrict our discussion to S-polarization in
which the ${\bf E}$-waves are nearly perpendicular to the sample
surface.

Using the Maxwell equation
\[
{\bf H}=\frac{1}{\omega\mu_0} {\rm rot} {\bf E}
\]
($\mu_0$ is the vacuum permeability, $\omega$ is the frequency of
the primary x-ray radiation) we obtain the following expression
for the Poynting vector
\beq{2}
{\bf S}=S_0 {\bf t}_0+S_{\bf h} {\bf t}_{\bf h},
\eeq
where ${\bf t}_{0,{\bf h}}$ are the unit vectors parallel to the
in-plane components of the wave vectors of the primary and
diffracted waves, respectively, and
\begin{eqnarray}
%
%
S_0(\phi,Z) &= &\frac{K}{2\mu_0\omega} {\rm Re} \left[
\sum_{n,m=1}^4 \E^{\I Z(k_{0nz}-k_{0mz}^*)} E_{0n}E_{0m}^* (1+\right . \nonumber \\
& & \left . c_n \E^{\I \phi})\right]\nonumber
\\
S_{\bf h}(\phi,Z) &=&\frac{K}{2\mu_0\omega} {\rm Re} \left[
\sum_{n,m=1}^4 \E^{\I Z(k_{0nz}-k_{0mz}^*)} E_{0n}E_{0m}^* c_m^*
(1+\right . \nonumber \\
& & \left . c_n \E^{-\I \phi})\right]
\label{e4}
\end{eqnarray}
are the components of the Poynting vector parallel to the primary
and diffracted beams, respectively. Since the amplitudes $E_{0n}$
and $E_{{\bf h}n}$ depend on the depth $Z$ of the Mn atom below
the surface, the entire Poynting vector depends on $Z$ as well. In
Eqs. (\ref{e4}), $\phi={\bf h}.{\bf r}$, where ${\bf r}$ is the
position of the Mn atom within the GaAs unit cell. Parameters
$E_{0n}$, $c_n$, and $k_{0nz}$ were calculated from the two-beam
dynamical diffraction theory taking into account four tie-points
on the dispersion surface.\cite{Authier:2001_a,Pietsch:2003_a}

The fluorescence signal is proportional to the weighted sum of the
contributions of all Mn atoms
\beq{5}
\Phi={\rm const}\int_{-\infty}^0 \D Z  \E^{-\mu_f |Z|}\sum_{p=}
n_p(Z) |{\bf S}(\phi_p,Z)|,
\eeq
where $\mu_f$ is the linear absorption coefficient of the
MnK$\alpha$ fluorescence radiation in GaAs and $n_p(Z)$ is the
concentration profile of the particular impurity in the (Ga,Mn)As
epilayer normalized to 1 ($p={\rm s,i^{(1)},i^{(2)}}$), e.g.,
$n_s$=1 for substitutional Mn occupying all Ga sites. Equation
(\ref{e5}) is the basis for numerical simulations; the resulting
fluorescence signal $\Phi$ is convoluted with the angular
distributions of the primary beam intensity both in azimuthal
direction and in the direction perpendicular to the sample
surface.  The fluorescence signal for randomly distributed Mn
atoms, Mn$_{\rm r}$, is simulated numerically by averaging the
value of $\Phi$ obtained from Eq. (\ref{e5}) over all values of
the phase $\phi$. Note that Mn$_{\rm r}$ atoms can represent
another phases present in the sample (MnAs for instance), or can
be incorporated, e.g., in an amorphous surface oxide layer.

\begin{figure}[t]
\epsfig{width=6cm,file=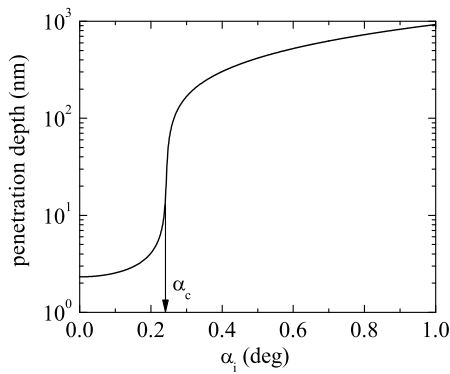}
\caption{The dependence of the penetration depth of the primary
x-ray beam on the incidence angle $\alpha_i$ in GaAs calculated
the energy of 10~keV; $\alpha_c$ denotes the critical angle of
total external reflection.}
\label{Figure3}
\end{figure}

An important advantage of the grazing-incidence geometry is that
the penetration depth of the primary x-ray radiation can be
sensitively tuned by changing its incidence angle $\alpha_i$, as
shown in Fig.~\ref{Figure3}. Slightly below the critical angle
$\alpha_c$ ($\alpha_c\approx 0.24^{\circ}$ in GaAs for 10~keV beam energy)
the penetration depth is only a few nanometers so that
the fluorescence signal stems from Mn atoms in a very thin surface
layer. When crossing the critical angle the penetration depth
steeply increases and for $\alpha_i$ only slightly larger than
$\alpha_c$ it exceeds 100 ~nm which is the thickness of the
studied (Ga,Mn)As layer.

\begin{figure}
\epsfig{width=8cm,file=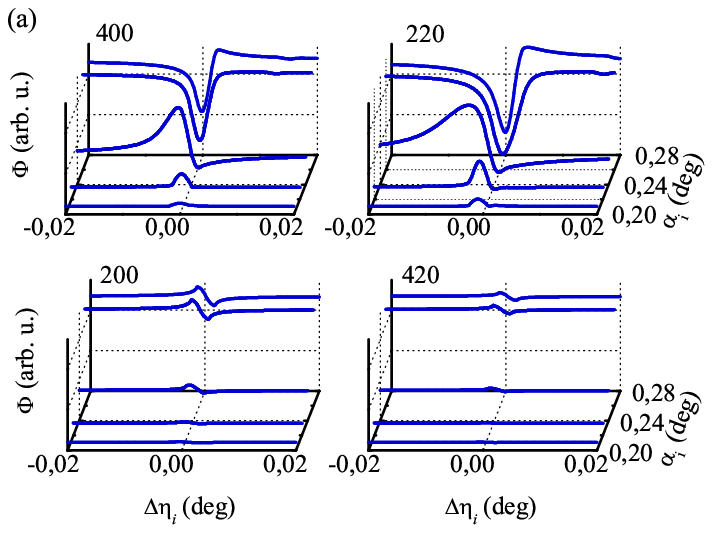}
\epsfig{width=8cm,file=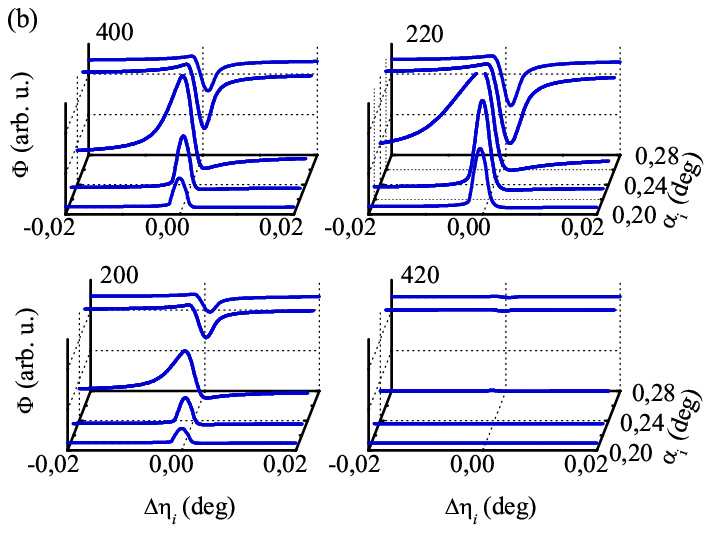}
\epsfig{width=8cm,file=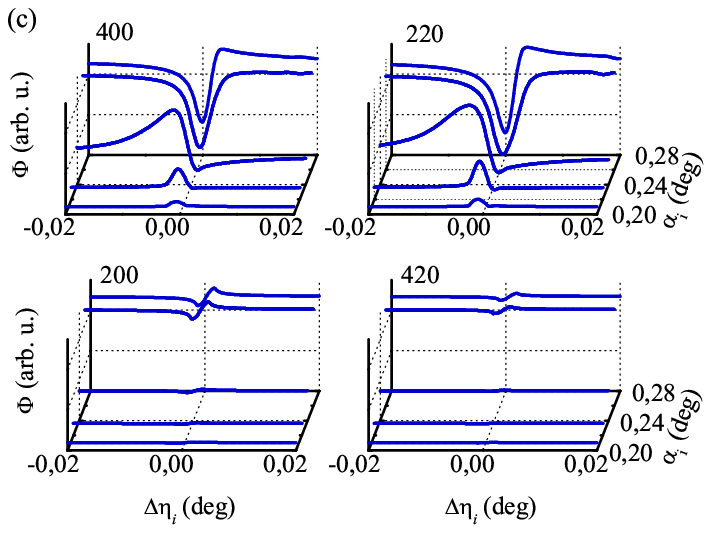}
\caption{The MnK$\alpha$ fluorescence intensities calculated for
the Mn atoms in substitutional positions (or in Mn$_{\I}^{(2)}$)
(a), in the random positions Mn$_{\rm r}$ (b), and in the
Mn$_{\I}^{(1)}$ positions (c), in four in-plane diffractions 400,
220, 200 and 420 (surface (001)).}
\label{Figure4}
\end{figure}

Fig.~\ref{Figure4} shows the fluorescence signal calculated for
substitutional Mn$_{\rm Ga}$, for interstitial Mn$_{\I}^{(1)}$,
and for  Mn$_{\rm r}$ positions. Note that for the (001) surface
orientation, Mn$_{\I}^{(2)}$ have the same lateral position as
Mn$_{\rm Ga}$ and, therefore, these two impurities cannot be
distinguished by measuring XSW fluorescence in the grazing
incidence geometry. The theoretical fluorescence signal is plotted
as a function of the azimuthal deviations $\delta \eta_i$ of the
primary x-ray beam from the diffraction maximum for several
incidence angles $\alpha_i=[0.20,0.22,0.24,0.26,0.28]$~deg near
the critical angle. The calculations were performed for in-plane
400, 220, 200 and 420 diffractions, assuming that the Mn atoms are
homogeneously distributed over the whole width of the (Ga,Mn)As
layer. The energy of the primary beam considered in the
calculations is 10~keV. The theory curves illustrate that
substitutional and random positions can be identified mainly from
the strong 400 and 220 diffractions while the Mn$_{\I}^{(1)}$
positions  from the weak 200 or 420 diffractions.

\begin{figure}
\epsfig{width=6cm,file=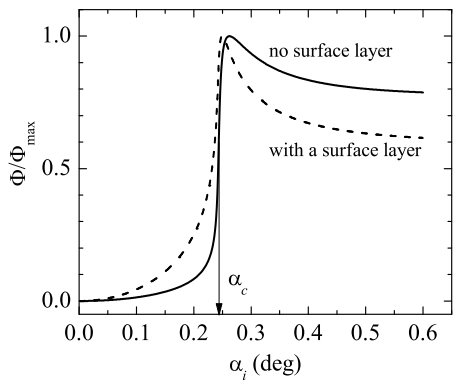}
\caption{The dependence of the fluorescence signal on the
incidence angle $\alpha_i$ calculated for the azimuthal direction
of the primary beam far from the diffraction position, assuming a
homogeneous distribution of Mn atoms (solid line), and a surface layer 10~nm
thick, where the Mn density is 10 times larger (dashed line). The curves do not
depend on the positions of the Mn atoms in the unit cell and on
the diffraction used. $\alpha_c$ is the critical angle of total
external reflection; the curves are normalized to their maxima.}
\label{Figure5}
\end{figure}

In Fig.~\ref{Figure5} we plot the calculated dependence of the
fluorescence signal on the incidence angle $\alpha_i$ far from the
diffraction position. Since the  signal here is excited by the
transmitted wave, its $\alpha_i$-dependence is determined only by
the penetration depth of the incoming radiation. Indeed, below the
critical angle the calculated fluorescence is very week while it
steeply increases for $\alpha_i>\alpha_c$. Note that the shape of
the curve does not depend on the position of Mn atoms in the unit
cell. To illustrate the sensitivity of the $\alpha_i$-dependence
of the fluorescence on inhomogeneities in the Mn depth profile,
two curves are plotted in Fig.~\ref{Figure5}; the solid line
corresponds to a uniform Mn distribution within the whole 100~nm
thick (Ga,Mn)As layer, the dashed line was obtained assuming a
10~nm Mn-rich surface layer with a 10$\times$ larger Mn density.


\section{Results}
\label{results}

We now present and analyze experimental data measured in the four
(Ga,Mn)As layers. Since the collected fluorescence data provide a
more direct information on Mn positions in the lattice and allow
us to study both the ordered epilayer part of the (Ga,Mn)As film
and the more disordered surface layer we start with this
technique. Results of the XSW experiments are also used to make a
more detailed interpretation of the XRD measurements of the
(Ga,Mn)As epilayer which are presented in the following
subsection. As will become clear by the end of this section the
two techniques, however, are largely complementary and  both
essential for the conclusions we arrived at in this work.

\subsection{X-ray standing-wave fluorescence from Mn atoms}
\label{XSW}

\begin{figure}
\epsfig{width=6cm,file=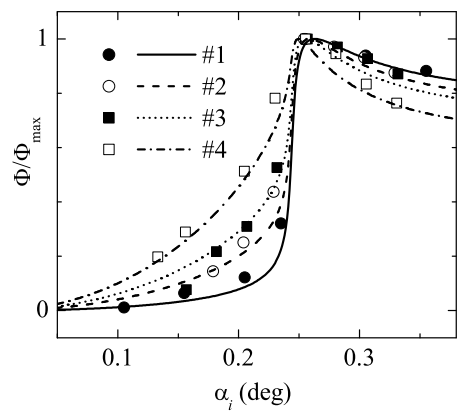}
\caption{The fluorescence signal measured for large angular
deviations $\delta\eta_i$ from the diffraction maximum for various
incidence angles (points) and its theoretical fit (lines). The
curves are normalized to their maximum values.}
\label{Figure6}
\end{figure}

Before resorting to the more elaborate standing-wave experiments,
the role of the surface during annealing can be tested by
fluorescence measurements without the XSW effect, i.e., for large
azimuthal deviations $\delta \eta_i$ from the diffraction maximum.
The comparison of experimental data shown in Fig.~\ref{Figure6}
with theoretical curves in Fig.~\ref{Figure5} clearly indicates
the development of a Mn-rich surface layer during the annealing
process. The experimental curves could be equally well fitted
assuming the thickness of the Mn-rich layer between  2 and 10~nm;
the fitted surface density then increases correspondingly to the
decrease of the assumed surface layer thickness.  This ambiguity
partly hinders quantitative interpretation, nevertheless, the
Mn-rich surface layer in annealed samples is ubiquitous. As an
illustration we give in Tab. \ref{t2} values of the fitted surface
Mn density $N$ relative to the density $n$ in the homogeneous
epilayer beneath the surface layer assuming a fixed Mn-rich layer
thickness of 3~nm for all four samples.

\begin{table*}
\begin{tabular}{c||cccc|cccc}
sample & $N/n$ & $p_{\rm r}$ & $p_{\I}^{(1)}$ & $P_{\rm r}$ &
$q_{\rm s}+q_{\I}^{(2)}$ & $q_{\rm r}$ & $Q_{\rm r}$\\
\hline
\#1 & $1.1 \pm 0.1$ & $0.12 \pm 0.01$ & $0.04 \pm 0.01$ & $2 \pm 1$ & 0.865& 0.1 & 0.62 \\
\#2 & $5 \pm 2$ & $0.22 \pm 0.01$ & $0.025 \pm 0.05$ & $5 \pm 1$ & 0.80 & 0.18 & 0.825 \\
\#3 & $12 \pm 3$ & $0.28 \pm 0.05$ & $0.035 \pm 0.01$ & $8 \pm 2$ & 0.76 & 0.21 & 0.88\\
\#4 & $17 \pm 3$ & $0.36 \pm 0.10$ & $0.06 \pm 0.01$ & $15 \pm 4$
& 0.705 & 0.255  & 0.935
\end{tabular}
\caption{The parameters obtained from the measured XSW curves.
$N/n$ is the ratio of the total Mn density in the surface layer
and the total Mn density $n=3.5\%$ in the homogeneous epilayer
beneath the surface layer. $p_{\rm r}$, $p_{\I}^{(1)}$ are the
densities of Mn$_{\rm r}$ and Mn$_{\I}^{(1)}$, respectively,
relative to the density of Mn$_{\rm Ga}$ plus Mn$_{\I}^{(2)}$
atoms, obtained from the XSW data at $\alpha_i>\alpha_c$. $P_{\rm
r}$ is the relative density of Mn$_{\rm r}$ obtained for
$\alpha_i<\alpha_c$. $q_{\rm s}+q_{\I}^{(2)}$ and $q_{\rm r}$ are
the recalculated partial concentrations of Mn$_{\rm Ga}$ plus
Mn$_{\I}^{(2)}$ and of Mn$_{\rm r}$, respectively, for
$\alpha_i>\alpha_c$; $Q_{\rm r}$ is the recalculated partial
concentration of Mn$_{\rm r}$ for $\alpha_i<\alpha_c$ (see the
text for more details).}
\label{t2}
\end{table*}

\begin{figure}
\epsfig{width=8cm,file=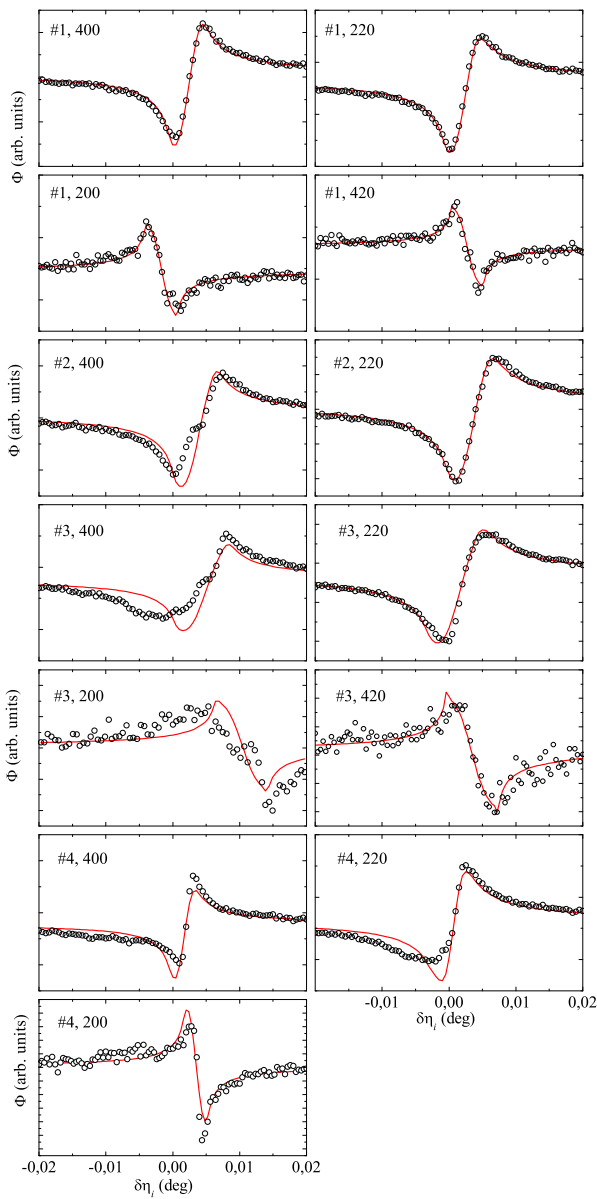}
\caption{The XSW scans of samples \#1 to \#4 measured in various
in-plane diffractions for the incidence angle $\alpha_i \approx
0.35$~deg ($>\alpha_c$) (points) and their theoretical fits
(lines).}
\label{Figure7}
\end{figure}

\begin{figure}
\epsfig{width=8cm,file=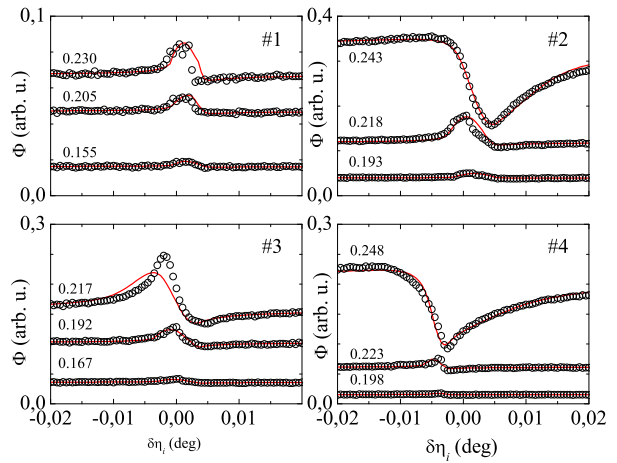}
\caption{The XSW scans of samples \#1 to \#4 measured in the 220
in-plane diffraction for three incidence angle smaller or close to
the critical angle $\alpha_c$ (points) and their theoretical fits
(lines). The values of the incidence angle $\alpha_i$ in deg are
indicated above the corresponding curve.}
\label{Figure8}
\end{figure}

The origin of Mn atoms arriving at the surface during annealing
and their incorporation in the surface layer is elucidated by the
fluorescence measurements in the grazing-incidence XSW geometry.
Examples of the measured $\Phi(\delta \eta_i)$ curves for
$\alpha_i>\alpha_c$ and $\alpha_i<\alpha_c$ are shown in Figs.
\ref{Figure7} and \ref{Figure8}, respectively.
For the incidence angle $\alpha_i\approx 0.35^{\circ}>\alpha_c$ the penetration
depth of the primary beam is much larger than 100~nm, i.e., the
XSW signal is averaged over the entire layer volume. The XSW
curves in Fig.~\ref{Figure7} were fitted for all diffractions simultaneously and from
the fit we could determine the density $p_{\rm r}$ of Mn$_{\rm r}$
relative to the density of the substitutional Mn$_{\rm Ga}$ atoms
plus Mn$_{\I}^{(2)}$ atoms, i.e., $p_{\rm r}=\bar{n}_{\rm r}/(\bar{n}_{\rm
s}+\bar{n}_{\I}^{(2)})$, and the relative density of Mn$_{\I}^{(1)}$,
$p_{\I}^{(1)}=\bar{n}_{\I}^{(1)}/(\bar{n}_{\rm s}+\bar{n}_{\I}^{(2)})$. Because of
the large penetration depth of the incidence beam the quantities
represent average densities over the entire (Ga,Mn)As layer.

Similar fitting procedure was applied to data measured at $\alpha_i<\alpha_c$, shown in
Fig.~\ref{Figure8}. Here the penetration depth of the primary beam
is only 3~nm so the fitted densities correspond to a thin surface
layer. The Mn$_{\rm r}$ atoms with high surface density were
detected even at this small incidence angle, however, the data
were not sensitive enough to measure the Mn$_{\I}^{(1)}$
impurities. For the fitting we used the values for the relative
Mn$_{\I}^{(1)}$ densities from the measurements at
$\alpha_i>\alpha_c$ and determined only the relative density
$P_{\rm r}$ of Mn$_{\rm r}$ in the surface layer. 

Results of the XSW experiments are summarized in Tab.~\ref{t2}.
For convenience we also included in the table partial Mn impurity
concentration recalculated relatively to the total combined density of
Mn$_{\rm Ga}$, Mn$_{\I}^{(2)}$, Mn$_{\I}^{(1)}$, and Mn$_{\rm r}$,
for both $\alpha_i<\alpha_c$ and $\alpha_i>\alpha_c$. In
particular, we show the values of the bulk relative densities
\[
q_{\rm r}=\frac{\bar{n}_{\rm r}}{\bar{n}}=\frac{p_{\rm r}}{1+p_{\rm
r}+p_{\I}^{(1)}},\ q_{\rm s}+q_{\I}^{(2)}=\frac{\bar{n}_{\rm s}+\bar{n}_{\I}^{(2)}}{\bar{n}}=\frac{1}{1+p_{\rm
r}+p_{\I}^{(1)}},
\]
 and
the relative density of the Mn$_{\rm r}$ atoms in the surface
layer
\[
Q_{\rm r}=\frac{P_{\rm r}}{1+P_{\rm r}+p_{\I}^{(1)}}.
\]
From the data in Tab.~\ref{t2} we can draw the following
conclusions: (i) The number of Mn$_{\rm r}$ atoms systematically
increases with annealing time. (ii) Since $P_{\rm r}/p_{\rm r}$ is
similar to the ratio between the (Ga,Mn)As layer thickness and the
exposed surface layer width in the $\alpha_i<\alpha_c$ experiment,
Mn$_{\rm r}$ atoms are concentrated in the surface layer. This
also means a marginal role (if any) of second phases in the
(Ga,Mn)As epilayer. (iii) As the Mn$_{\rm r}$ atoms fill the
surface, the density of Mn$_{\rm Ga}$ plus  Mn$_{\I}^{(2)}$
impurities in the (Ga,Mn)As layer decreases. The XRD data
discussed in the following subsection show that the density of
substitutional  Mn$_{\rm Ga}$ atoms remains constant during
annealing. We can, therefore, directly link the outdiffusion of
interstitial Mn$_{\I}^{(2)}$ from the (Ga,Mn)As with the build-up
of the Mn$_{\rm r}$-rich surface layer. Note that we detected only
a small density of Mn$_{\I}^{(1)}$ impurities and no systematic
behavior during annealing can be extracted from the data.

\begin{figure}
\epsfig{width=8cm,file=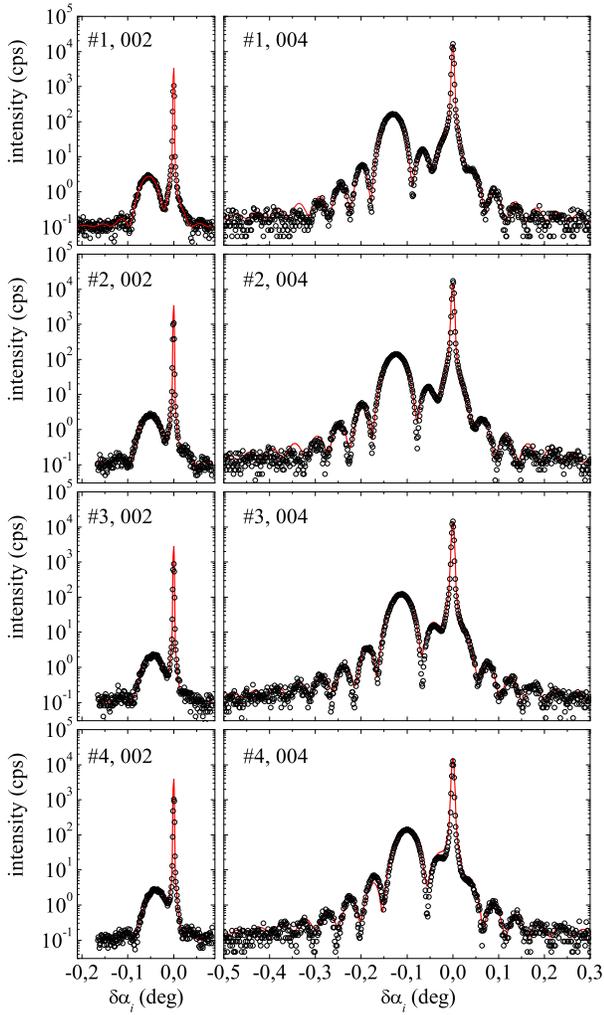}
\caption{High-resolution x-ray diffraction curves of samples \#1
to \#4 in diffractions 002 (left column) and 004 (right column)
(points) and their fits by conventional dynamical diffraction
theory (lines).}
\label{Figure9}
\end{figure}

\subsection{X-ray diffraction from (Ga,Mn)As epilayers}
\label{XRD}

Experimental XRD curves, along with their fits obtained by
conventional dynamical diffraction theory, are shown in
Fig.~\ref{Figure9}. The 002 and 004 diffractions were fitted
simultaneously. Local distortions of the lattice around Mn$_{\rm
s}$ and Mn$_{\I}^{(1,2)}$ atoms were neglected in the fitting
procedure. The very good correspondence of experimental and
theoretical diffraction curves demonstrates good structural
quality of the epilayers and homogeneity of the Mn distribution.
No measurable Mn concentration gradients were detected across the
epilayer.

The epilayer thickness and the total lattice expansion parameter
$\xi$, defined as $a=a_0+\xi$ where $a_0$ is the GaAs lattice
constant and $a$ is the relaxed lattice parameter of the (Ga,Mn)As
epilayer, were obtained independently with error bars smaller than
1\%. Since the average structure factor and consequently the
heights of the layer maxima on the diffraction curves depend on
the impurity type, we were able to measure lattice expansion
contributions from individual impurities, albeit with smaller
accuracy. In particular,  we assumed that $\xi=\xi_{\rm
s}+\xi_{\I}+\xi_{\rm a}$, where the subscript "s" stands for
substitutional Mn$_{\rm Ga}$, "i" for interstitial Mn of both
types Mn$_{\I}^{(1)}$ and Mn$_{\I}^{(2)}$, and we also included
antisite As$_{\rm Ga}$ impurities ($\xi_a$). Note that the
expansion parameters $\xi_p, p={\rm s,i,a}$ and the Vegard's law
dilatation coefficients $\beta_p$ are related as,
$\xi_p=\beta_pn_p$.

The influence of Mn$_{\rm r}$ atoms on $\xi$ was not taken into
account in our simulations. Other crystallographic phases, where
Mn$_{\rm r}$ atoms are incorporated, could affect the mean lattice
constant of the epilayer only if they deform the surrounding
crystal lattice and give rise to measurable diffuse x-ray
scattering. Consistent with the results of XSW experiments, no
such effects were detected in the XRD data for the studied
(Ga,Mn)As epilayers.

We found that equally good fits could be obtained assuming any
value of $\xi_{\rm a}$ ranging from 0 to $5\times 10^{-3}$~\AA.
The extracted total expansion $\xi$, shown in Tab.~\ref{t1}, is
independent of the choice of $\xi_{\rm a}$ which immediately
implies the dependence of the fitted $\xi_s$ and $\xi_i$ on
$\xi_{\rm a}$. In Tab.~\ref{t1} we show expansion parameters
obtained for the two limiting values of $\xi_{\rm a}$. Despite the
quantitative ambiguities in the values of individual expansion
parameters we can safely conclude that the lattice constant
decreases during annealing, that the density of Mn$_{\rm Ga}$
atoms remains constant and, consequently, the decrease of the
total lattice parameter with annealing time is due to the removal
of Mn$_{\I}$ impurities from the ordered epilayer part of the
(Ga,Mn)As film. (Note that As$_{\rm Ga}$ antisites in GaAs are
stable up to $\sim 450^{\circ}$C.\cite{Bliss:1992_a})

Although the XRD and XSW experiments did not allow to separately
determine Mn$_{\rm Ga}$ and Mn$_{\I}$ partial densities, the
p-type character of the (Ga,Mn)As semiconductor and the single
acceptor nature of Mn$_{\rm Ga}$ and the double-donor nature of
Mn$_{\I}$ and of As$_{\rm Ga}$ imply that  $n_{\rm s}>2n_{\I}$.
From this and from the expansion parameters in Tab.~\ref{t1}  we
obtain that the dilatation coefficient $\beta_{\I}$ is at least
10$\times$ larger than  $\beta_{\rm s}$. A significantly larger
lattice expansion due to Mn$_{\I}$ as compared to Mn$_{\rm Ga}$ is
consistent with \emph{ab initio} theory
predictions.\cite{Masek:2003_a,Masek:2005_a}

Using the results of XSW measurements and recalling that Mn$_{\I}$
are the only unstable impurities during annealing which contribute
to $\xi$ we can also make an estimate for the absolute value of
$\beta_{\I}$. The measured changes in $\xi$ after each annealing
step and the Mn$_{\I}$ expansion coefficient are related as,
$\beta_{\I}=\Delta\xi_{\I}/\Delta n_{\I}$, where $\Delta
n_{\I}=n[\Delta q_{\I}^{(1)}+\Delta(q_{\rm s}+q_{\I}^{(2)})]$. For
$n$ we can take the MBE growth value $3.5\pm 0.5$\%, and the
changes of the relative partial concentrations can be read out
from Tab~\ref{t2}. The value $\beta_{\I}=0.4\pm 0.2$ we arrive at
has an appreciable scatter, reminiscent of the current theoretical
uncertainty for this coefficient.\cite{Masek:2003_a,Masek:2005_a}
Nevertheless, the experiment and theory are consistent within
these error bars.

\begin{table*}
\begin{tabular}{c||c|cc|cc}
sample & $\xi=\sum_{p={\rm s,i,a}} \beta_p n_p$ ($10^{-3}$\AA) &
$(\beta_{\rm s} n_{\rm s})$ ($10^{-3}$\AA) & $(\beta_{\I} n_{\I})$
($10^{-3}$\AA) & $(\beta_{\rm s}n_{\rm s})$ ($10^{-3}$\AA)
& $(\beta_{\I} n_{\I})$ ($10^{-3}$\AA)\\
&  & \multicolumn{2}{|c|}{for $\xi_{\rm a}=0$} &
\multicolumn{2}{|c}{for $\xi_{\rm a}=5\times 10^{-3}$~\AA}\\
\hline
\#1 & $10.2 \pm 0.1$ & $1.3 \pm 0.6$ & $8.9
\pm 0.5$  & $0.9 \pm 0.6$ & $4.3 \pm 0.5$ \\
\#2 & $9.6 \pm 0.1$ & $1.4 \pm 0.6$ & $8.2
\pm 0.5$  & $0.8 \pm 0.6$ & $4.0 \pm 0.5$\\
\#3 & $8.9 \pm 0.1$ & $1.3 \pm 0.6$ & $7.6
\pm 0.5$  & $0.8 \pm 0.6$ & $3.1 \pm 0.5$\\
\#4 & $8.0 \pm 0.1$ & $1.4 \pm 0.6$ & $6.6
\pm 0.5$ & $0.9 \pm 0.6$ & $2.1 \pm 0.5$\\
\end{tabular}
\caption{Lattice expansion parameters of the (Ga,Mn)As epilayers
determined from the XRD measurements.}
\label{t1}
\end{table*}

\section{Discussion}
\label{discussion}

Our experiments confirm the basic picture derived from previous
studies of Mn incorporation in (Ga,Mn)As and of the annealing
process but the measurements are more direct and provide more
detailed information about the structure of the (Ga,Mn)As epilayer and of the layer surface.
The XSW measurements clearly indicate the existence of a Mn$_{\rm r}$-rich
surface layer. The detailed chemistry of this layer is unknown to us; it
could be a surface oxide growing during annealing of (Ga,Mn)As in
air, or a very thin MnAs layer. 

The small number of Mn$_{\I}^{(1)}$ we detected also requires
further investigation. In particular, it would be useful to use
a XSW geometry in which both Mn$_{\I}^{(1)}$ and
Mn$_{\I}^{(2)}$ can be detected independently. Such a geometry together with 
calibration measurements on pure Mn crystal should allow for a
fully quantitative interpretation of the XSW data. In order to
detect Mn$_{\I}^{(2)}$ independently of Mn$_{\rm Ga}$ atoms, it would be necessary to use a
diffraction with a non-zero out-of-plane component of the diffraction vector
${\bf h}$, since Mn$_{\rm Ga}$ and Mn$_{\I}^{(2)}$ atoms differ
only in their vertical positions. For such a diffraction, the XSW
effect can be measured (i) in the case of a very thin (Ga,Mn)As layer
or (ii) for a very thick layer. In the former case, the
diffraction in substrate can be used for the creation of the
standing wave and the (Ga,Mn)As layer has to be very thin so that the
vertical misalignment of the Mn atoms (due to the lattice mismatch)
with respect to the standing wave pattern is negligible. This
condition is fulfilled for the layer thickness
\beq{10}
T \ll \frac{a_{\rm sub}}{f},
\eeq
where $f=(a_{{\rm layer}\bot}-a_{\rm sub})/a_{\rm sub}$ is the
vertical lattice mismatch of the (Ga,Mn)As epilayer with respect to the GaAs substrate.
Since, in our case, the mismatch is of the order of 10$^{-2}$, the
(Ga,Mn)As layer would have to be much thinner than 50~nm.

In the case (ii), the diffraction in the (Ga,Mn)As epilayer can be used as a source
of the standing wave. In order to achieve a measurable XSW effect,
the intensities of the primary and diffracted beams must be
comparable, i.e., the thickness of the epilayer has to be comparable
to the x-ray extinction length (above 1~$\mu$m). Note however, that theoretical
analysis of these experiments will be complicated by the fact that the
standing wave itself is affected by the presence of Mn atoms.

\section{Conclusions}
\label{conclusions}

We have studied the structure of epitaxial (Ga,Mn)As layers and their
changes during post-growth annealing, using high-resolution
x-ray diffraction and grazing-incidence x-ray standing-wave  
techniques. The layers are formed by a high crystal quality (Ga,Mn)As epilayers
covered by a thin Mn-rich layer. We have identified that the excess surface Mn density is
due to Mn atoms occupying random positions incommensurate with the host
lattice. The increase of the density of random Mn atoms during annealing
is correlated with the decrease of interstitial Mn density in the (Ga,Mn)As epilayer.

\section*{Acknowledgment}

We acknowledge fruitful collaborations with Dr. Tien-Lin Lee and
other members of the beamline ID32 team at ESRF Grenoble, and
support from the Ministry of Education of the Czech Republic under
Grants No. MSM 0021620834 and LC510, from the Grant Agency of
Czech Republic under Grants No. 202/06/0025, 202/05/0575, and
202/04/1519, and from the UK EPSRC under Grant No. GR/S81407/01.


\end{document}